\documentclass[A5paper,useAMS,usenatbib,usegraphicx]{mn2e}

\title[]{Characteristics of the Galaxy according to Cepheids}
\author[Majaess, Turner and Lane]
{D.~J. Majaess$^{1,2}$\thanks{Email: dmajaess@ap.smu.ca}, D.~G. Turner$^{1}$\thanks{Email: turner@ap.smu.ca} and D.~J. Lane$^{1,2}$\thanks{Email: dlane@ap.smu.ca} \\ \\
  $^1$Department of Astronomy and Physics, Saint Mary's University,
    Halifax, Nova Scotia B3H 3C3, Canada \\
  $^2$The Abbey Ridge Observatory, Stillwater Lake, Nova Scotia, Canada
}
\date{Accepted 2009 Month xx. Received 2009 Month xx}
\pagerange{\pageref{firstpage}--\pageref{lastpage}} \pubyear{2009}

\begin{document}

\label{firstpage}

\maketitle

\begin{abstract}
Classical and Type II Cepheids are used to reinvestigate specific properties of the Galaxy. A new Type II reddening-free Cepheid distance parameterization is formulated from LMC Cepheids (OGLE), with uncertainties typically no larger than 5--15\%. A distance to the Galactic centre of $R_0 = 7.8\pm0.6$ kpc is derived from the median distance to Type II Cepheids in the bulge (OGLE), $R_0 = 7.7\pm0.7$ kpc from a distance to the near side of the bulge combined with an estimated bulge radius of $1.3\pm0.3$ kpc derived from planetary nebulae.  The distance of the Sun from the Galactic plane inferred from classical Cepheid variables is $Z_{\sun}=26\pm3$ pc, a result dependent on the sample's distance and direction because of the complicating effects of Gould's Belt and warping in the Galactic disk. Classical Cepheids and young open clusters delineate consistent and obvious spiral features, although their characteristics do not match conventional pictures of the Galaxy's spiral pattern. The Sagittarius-Carina arm is confirmed as a major spiral arm that appears to originate from a different Galactic region than suggested previously. Furthermore, a major feature is observed to emanate from Cygnus-Vulpecula and may continue locally near the Sun.  Significant concerns related to the effects of metallicity on the {\it VI}-based reddening-free Cepheid distance relations used here are allayed by demonstrating that the computed distances to the Galactic centre, and to several globular clusters (M54, NGC 6441, M15, and M5) and galaxies (NGC 5128 and NGC 3198) which likely host Type II Cepheids: agree with literature results to within the uncertainties.  An additional empirical test is proposed to constrain any putative metallicity dependence of Cepheid distance determinations through forced matches of distance estimates to a particular galaxy using both Type II and classical Cepheids.
\end{abstract}


\section{Introduction}
The value of Cepheid variables as distance indicators is well established by their continued use as standard candles for the extragalactic distance scale \citep{ke99,fr01,th03,pi06,fe07,gi08}. That same property can also be used to map the Milky Way's spiral arms and to establish various fundamental parameters for the Galaxy, as pointed out previously \citep[e.g.,][]{ks63,fe68,cc87,op88,be06}. 

The present study capitilizes on recent advances in the field which enable the use of Type II Cepheids and classical Cepheids to place stronger constraints on specific properties of the Galaxy. With regards to Type II Cepheids, a new reddening-free distance relation is formulated here and calibrated using LMC Type II Cepheids discovered by OGLE \citep{ud99,so08}. In relation to classical Cepheids, the present study makes use of a new calibration of the reddening-free classical Cepheid distance relation by \citet{ma08a}, which is tied to established cluster Cepheids \citep[e.g.,][]{tu02} and new HST parallax measures \citep{be07}. The parameterization appears to be capable of reproducing classical Cepheid distances with uncertainties typically no larger than $\pm5\%$ to $\pm15\%$, where the larger value of the uncertainty takes into account extreme variations in location in the instability strip and the reddening law throughout the Galaxy \citep[see][]{tu89,tu96a}, given the reddening-free relationship is tied to a Galactic average. The results of \citet{ma01} also support the assumption of a standard reddening law, to first order, when determining the distances to extragalactic Cepheids.  Nevertheless, the relationship itself replicates known distances to Cepheid calibrators to within $\pm4\%$, and that includes Cepheids well distributed about the centre of the observational instability strip. Older relationships of comparable type are generally tied to calibrators whose parameters have since been revised.

The present study also utilizes a new and enlarged sample of classical Cepheid variables with multi-passband photoelectric and CCD photometry \citep[e.g.][]{sz77,sz80,sz81,sz83,be92,be94,be97,be98,be00}. In most cases precision photoelectric and CCD photometry enable the pulsation mode of a classical Cepheid to be constrained by means of Fourier analysis \citep{za05,bs98,we95,be95}, resulting in improved distance estimates for shorter period objects. Efforts to discover additional Cepheids through all-sky variability surveys also help to expand the Galactic sample, e.g., ASAS \citep{po00}, TASS \citep{dr06}, and NSVS \citep{wo04}. 

This paper is organised as follows. In \S \ref{gcsection} a Type II reddening-free Cepheid distance relation is developed and tested by determining the distance to the Galactic centre and several globular clusters and galaxies. \S \ref{bulgesection} tackles the thickness of the Galactic bulge by means of bulge planetary nebulae and an estimated distance to the Galactic centre.  \S \ref{szsection} uses classical Cepheids to determine the Sun's distance above the Galactic plane and to trace the warping of the Galactic disk. Lastly, \S \ref{spiralsection} uses classical Cepheids and young open clusters to delineate local Galactic spiral structure. 

\section{Distance to the Galactic Centre}
\label{gcsection}

Classical Cepheids currently provide only indirect information about the distance to the Galactic centre, through their kinematics. Yet abundant numbers of their low-mass Type II counterparts are detected in the Galactic bulge. Distances to Type II Cepheids can be established by first constructing a reddening-free distance relation like that derived for classical Cepheids \citep{ma08a}. The calibrators are LMC Type II Cepheids, with an adopted zero-point to the LMC established from classical Cepheids and other means \citep[$\sim18.50$,][]{ls94,fr01,be02,be07,vl07,fo07,ma08a}.  Although there are fellow research groups that propose the LMC is closer \citep{ud98}.  The distances were then computed for Type II Cepheids lying in the direction of the Galactic bulge.

The distance to a classical Cepheid can often be estimated fairly reliably via a reddening-free relation of the following form \citep{vb68,ma82,op83,ma08a}:
\begin{equation}
\label{eqn1}
5\log{d}=V+ \alpha \log{P} + \beta (V-I) + \gamma \;, 
\end{equation}
assumed here to be true for Type II Cepheids as well as classical Cepheids. A calibrating set of LMC Type II Cepheids from the OGLE survey \citep{ud99,so08} was used to determine the co-efficients of equation (\ref{eqn1}) that minimize the $\chi^2$ statistic, yielding the solution: 
\begin{equation}
\label{eqn2}
5\log{d}=V+ 2.34 \log{P} - 2.25 (V-I) + 6.03 + \phi \;.
\end{equation}

\begin{figure}
\includegraphics[width=8cm]{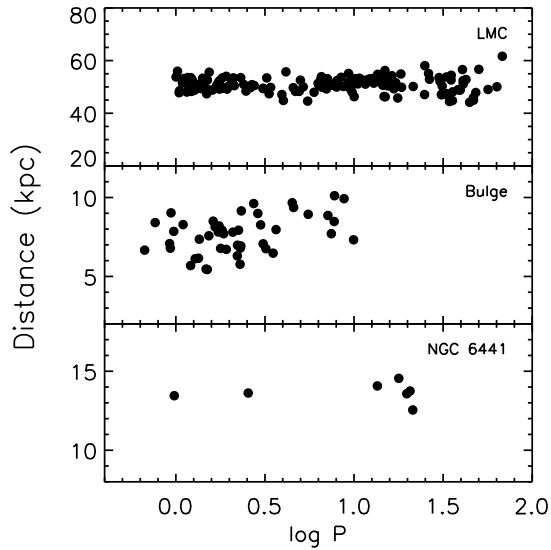}
\caption{The computed distances (equation \ref{eqn2}) to Type II Cepheids in the LMC (top, calibrating sample), the Galactic bulge (middle), and the globular cluster NGC 6441 (bottom).  The data are plotted as a function of pulsation period.}
\label{fig1}
\end{figure}

A plot of the computed distances to the calibrating set is shown in Fig. \ref{fig1}. The average deviation is $\sim5$\% and comparable to the uncertainties obtained by reddening-free classical Cepheid distance relations when reproducing calibrating data sets \citep{ma08a}. A correction term of $\phi=0.05\times|\log{P}|^{4.8}$ is adopted to linearize the equation over all period ranges from the BL Her to the RV Tau regimes, given that different classes of Type II Cepheids appear to be matched to different Wesenheit functions \citep{so08}. The above relationship yields reliable results for Type II Cepheids with periods of $\log{P} \le 1.6$, but is not calibrated for use beyond that limit. The correction term ($\phi$) can be updated when the necessary calibrators become available.

Distances to a selection of Type II Cepheids identified by OGLE as lying in the Galactic bulge \citep{ku03} were computed using equation (\ref{eqn2}), and are plotted in Fig. \ref{fig1}. The median distance to bulge Type II Cepheids analyzed via equation (\ref{eqn2}) implies a distance to the Galactic centre of $R_0 = 7.8\pm0.6$ kpc, with the caveat that the Type II Cepheids are assumed to be symmetrically distributed about the centre. A second estimate for the distance to the Galactic centre was established by adding an estimate for the radius of the bulge to the distance to the near side of the Galactic bulge as identified by Type II Cepheids, i.e. $R_0 = R_{\rm NS}+\beta$, under the assumption that the Galactic bulge is spherically symmetric. The situation is less simple if there is a central bar. The near side (NS) of the distribution is estimated to lie at a distance of $R_{\rm NS} = 6.4\pm0.4$ kpc, although admittedly, this value is dependent on whether the scatter in Fig. \ref{fig1} is inherent to the true distances of Type II Cepheids. A correction factor of $\beta = 1.3\pm0.3$ kpc was adopted from a geometric estimate for the radius of the bulge (see \S \ref{bulgesection}), giving a value of $R_0 = R_{\rm NS}+\beta=7.7\pm0.7$ kpc.

The data in Fig. \ref{fig1} indicate an apparent dependence of distance with pulsation period for bulge Type II Cepheids, but no such bias is noted for the distances computed to Type II Cepheids in the metal-rich globular cluster NGC 6441.  The observed trend for the bulge data may be a sampling effect, but there is also a possibility that it is tied to a metallicity dependence in the reddening-free Type II Cepheid distance parameterization.  Classical Cepheids in the LMC and their Galactic counterparts exhibit different metallicities \citep{lu98,an02,mo06}, and such differences probably extend to Type II Cepheids.  Yet the slope of a {\it VI} classical Cepheid relation is relatively unaffected by metallicity \citep{ud01,pi04,be07,vl07,fo07,ma08a}, and indeed, equation \ref{eqn2} is also a $VI$-based relation. \citet{ud01} and \citet{pi04} also suggest the zero-point of the classical Cepheid PL relation ($VI$) is insensitive to metallicity, although there are fellow research groups that propose a modest correction \citep[e.g.,][]{ke98,ma06,sc09}.  The current body of evidence appears to indicate that the effect of metallicity on {\it VI}-based classical Cepheid distance relations is small in comparison with other concerns and uncertainties, especially in relation to extragalactic observations. This notion likely extends to the {\it VI} reddening-free Type II Cepheid distance relation presented here, namely since the computed distances to the Galactic centre and to several globular clusters and galaxies by means of equation \ref{eqn2} agree with literature results to within the uncertainties (demonstrated below).  Ultimately, larger statistics are needed to explore and characterize any possible bias, especially \textit{vis \`a vis} the bulge data.

Recent studies by \cite{fe08} and \citet{gr08} established distances to the Galactic centre from Type II Cepheids and RR Lyrae variables of $7.64\pm0.21$ kpc and $7.94\pm0.37$ kpc, respectively, consistent with a geometric estimate of $7.94\pm0.42$ kpc obtained by \citet{ei03} from the orbital motion of star S2 about Sgr A*. The above values match the distances estimated here to the Galactic centre, and are consistent with similar values deduced from planetary nebulae in the Galactic bulge \citep[e.g.,][]{po90,re93}.

\begin{figure}
\begin{center}
\includegraphics[width=8cm]{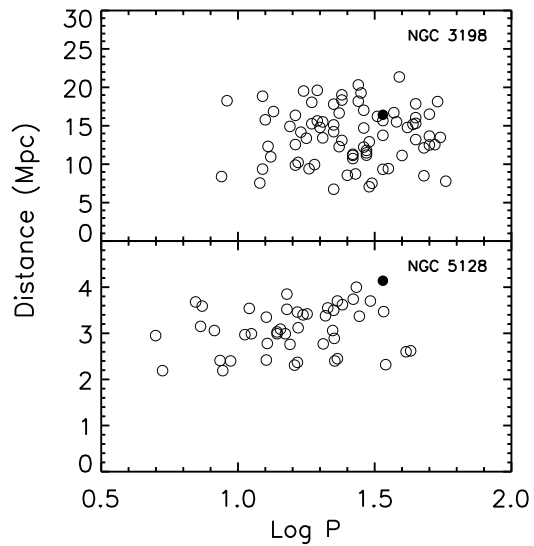}
\caption{Cepheid period-distance diagrams for the galaxies NGC 3198 (upper) and NGC 5128 (lower), with filled circles identifying stars analyzed using the Type II Cepheid distance relation, equation (\ref{eqn2}). Open circles identify stars analyzed with the classical Cepheid distance parameterization \citep{ma08a}.}
\label{fig2}
\end{center}
\end{figure}

Another test of the reliability of the {\it VI} reddening-free Type II Cepheid distance parameterization can be made using globular clusters. \citet{pr03} provide a convenient summary of the limited {\it VI} photometry available for Type II Cepheids in globular clusters (their Tables 7 and 8), which, in the absence of a larger data set, permits a comparison of distances computed to the clusters by equation (\ref{eqn2}) with literature results. The resulting distances derived for 10 Type II Cepheids in the globular clusters M54, M92, and NGC 6441 (see Fig. \ref{fig1}) agree with literature values for their distances, with the average difference, in the sense present--literature values, being $+5\pm4$\% (the datapoint for M92 is most deviant). A minor cautionary note is that the data for NGC 4372 given in Table 8 of \citet{pr03} are not mean magnitudes, and the stars require additional observations \citep[see][]{kk93}. Two variable stars discovered with Cepheid-like light curves in M15 are likely Type II Cepheids \citep[V1 \& V86,][]{co08}, leading to a distance of $11.1\pm0.8$ kpc (equation \ref{eqn2}).  This is consistent with the estimated distance of $10.4\pm0.8$ kpc to M15 \citep{dh93}. In addition, V42 and V84 in M5 \citep{ran07,rab07} are probably Type II Cepheids given their Cepheid-like light curves and computed distance of $d\sim7.5$ kpc (equation \ref{eqn2}), in agreement with the distance to M5 \citep[e.g.][]{la05}. The aforementioned globular clusters exhibit a large range in metallicity ($\Delta [Fe/H]\simeq1.75$, \citet{ha96}), so the close agreement of the present distance estimates with literature results negates a sizeable metallicity effect.  

An independent test is possible using galaxies, since the {\it VI} reddening-free Type II Cepheid distance relation (equation \ref{eqn2}) should provide reasonable distances for extragalactic Cepheids. A literature search was made with the assumption that Type II Cepheids will yield overly large distances when computed using a classical Cepheid distance relation. Two such instances were found: star C33 in NGC 3198 \citep{ke99} and star C43 in NGC 5128 \citep{fe07}. Both stars exhibit Cepheid-like light curves and were discovered from searches for classical Cepheids in the galaxies by those research teams. Cepheid period-distance diagrams in Fig. (\ref{fig2}) for both galaxies indicate that the two stars are probably Type II Cepheids and members of NGC 3198 and NGC 5128 respectively, once their distances are computed with the appropriate parameterization (equation \ref{eqn2}). The former object may be the most distant Type II Cepheid established to date, with an estimated distance of $d=13.7\pm3.6$ Mpc. Admittedly, the uncertainties are large, but such cases demonstrate the potential use of Type II Cepheids for extragalactic distance determinations. Type II Cepheids may also offer an empirical resolution to the metallicity question, given that, for a particular galaxy, distances computed from reddening-free classical and Type II Cepheids (equation \ref{eqn2}) should yield comparable results if metallicity effects are relatively small.

Finally, the location of Type II Cepheids detected in windows towards the Galactic bulge is somewhat irregular, although that does not appear to affect the present estimates for $R_0$. It may be advantageous in future studies to map the spatial location of sample members to outline the bulge distribution, as a means of eliminating potential sources of bias and of inferring the shape and inclination of the bulge \citep[e.g.,][]{ku03}. The spatial structure of the Magellanic Clouds has been successfully determined by similar means \citep{cc86,ls86,we87,ni04}. 

\begin{figure}
\begin{center}
\includegraphics[width=7cm]{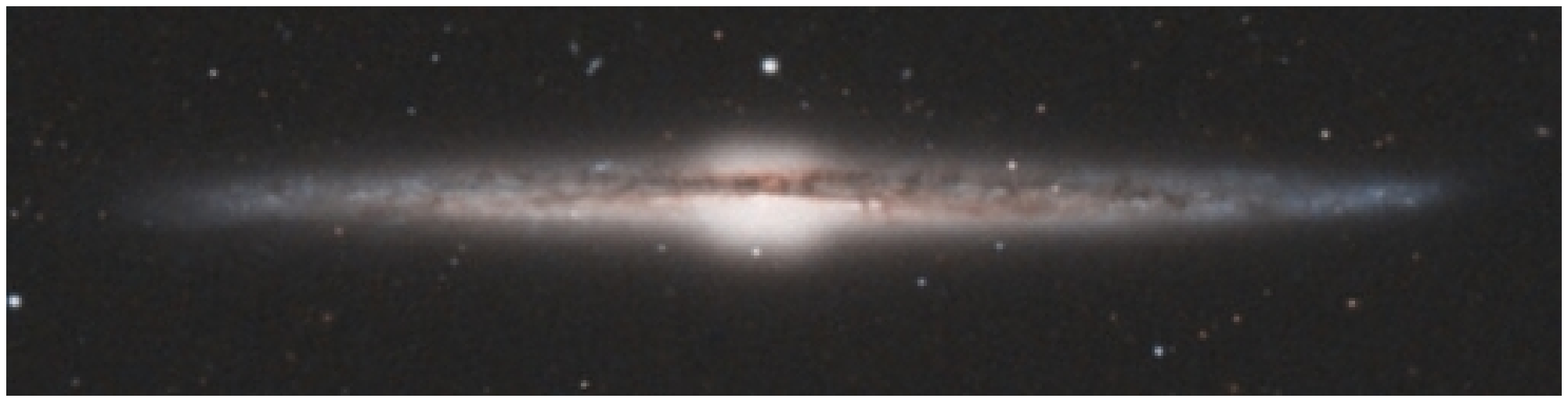}
\end{center}
\includegraphics[width=7.7cm]{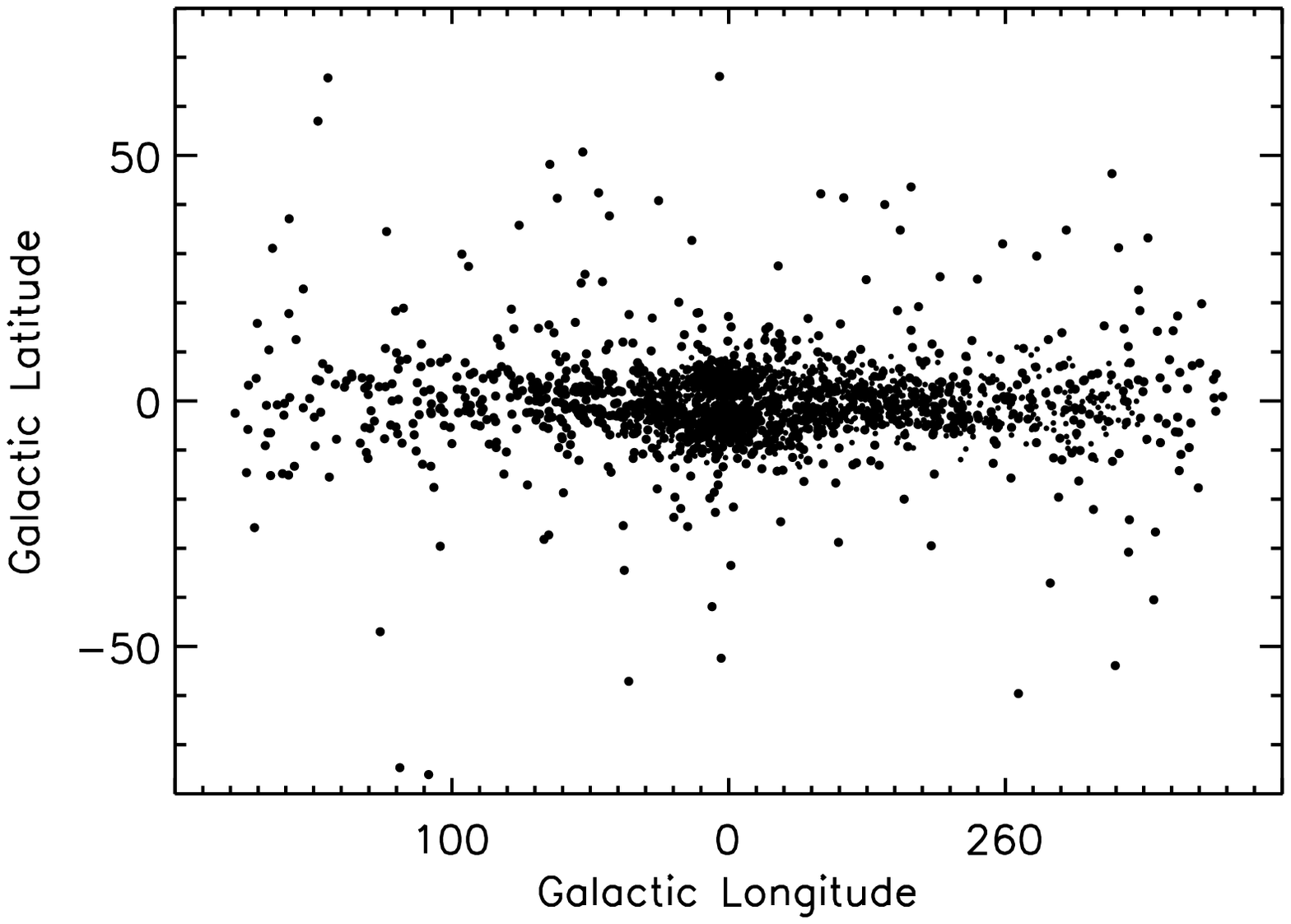}
\caption{A pseudo colour image of NGC 4565 (upper) constructed from POSS II data (Noel Carboni), and the distribution of planetary nebulae in Galactic co-ordinate space (lower), compiled from the catalogues of \citet{ko01} and MASH I \& II \citep{pa06,mi08}.}
\label{fig3}
\end{figure}

\section{Thickness of the Bulge}
\label{bulgesection}
The accepted view of the Galaxy's edge-on structure has been for many years that illustrated by \citet{pl27,pl36} and \citet{ga57}. Plaskett's envisioned structure agrees well with the distribution of planetary nebulae in Galactic co-ordinate space (Fig. \ref{fig3}), compiled from the catalogues of \citet{ko01} and MASH I \& II \citep{pa06,mi08}. Planetary nebulae, whose progenitors are primarily old, low-mass objects, outline the Galactic bulge, where their distribution peaks rather clearly \citep[see Fig. 1 of][]{ma07}. The maximum apparent thickness of the Galactic bulge perpendicular to the Galactic plane can be established from the co-ordinates of planetary nebulae in Fig. \ref{fig3}, which imply a bulge thickness in latitude of about $\pm9\degr$. From geometry and an estimated distance to the Galactic centre of $R_0 = 8\pm1$ kpc, a reasonably all-encompassing value \citep{re93}, the maximum apparent thickness of the bulge along $\ell \simeq 0 \degr$ is given by $H_{\rm B} = 2 \times R_0 \times \tan 9\degr = 2.5\pm0.3$ kpc. If the Galactic bulge is spherically symmetric, then the adopted value for $\beta$ (the radius of the bulge) in the previous analysis is justifed. A possible complication can be seen in Fig. \ref{fig3}, since bulge planetary nebulae appear to lie primarily below $b=0\degr$.

\section{The Sun's Distance from the Galactic Plane}
\label{szsection}
The distance to a classical Cepheid, $d$, can be approximated using the reddening-free equation given by \citet{ma08a}:
\begin{equation}
\label{eqn5}
5\log{d}=V+ (4.42) \log{P} - (3.43) (\langle B \rangle-\langle V \rangle) + 7.15 \;,
\end{equation}
where $P$ is the period of pulsation, and $\langle B \rangle$ and $\langle V \rangle$ are Johnson blue and visual mean magnitudes. Equation \ref{eqn5} is a useful formulation, given the increased availability of classical Cepheids with mean $BV$ photometry. A classical Cepheid's projected distance from the Galactic plane is found geometrically: $Z=d \times \sin{b}$, where $b$ is Galactic latitude, compiled for each classical Cepheid from the \textit{General Catalogue of Variable Stars} \citep{sd04}. 

\begin{figure}
\begin{center}
\includegraphics[width=8cm]{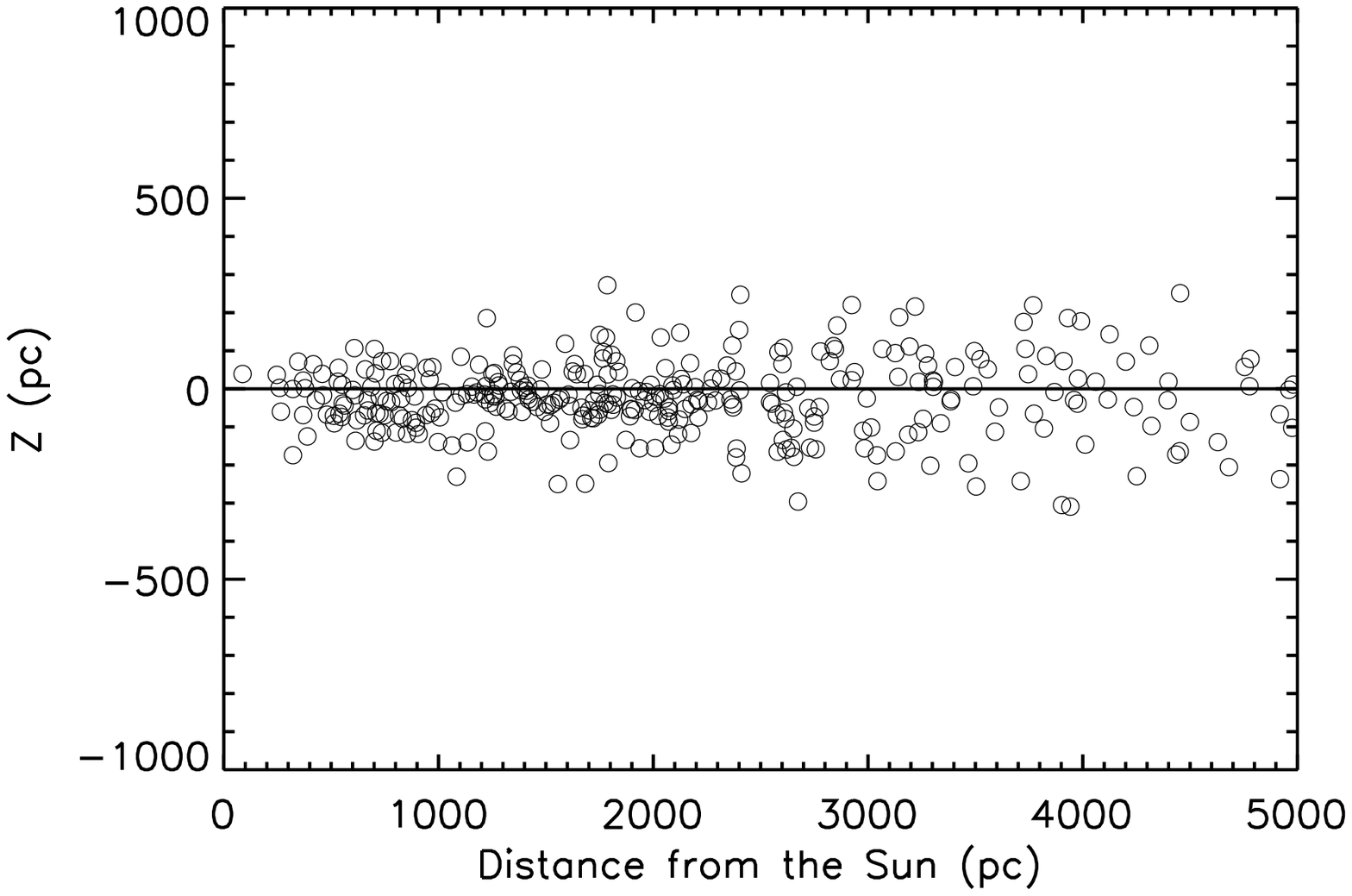}
\includegraphics[width=8cm]{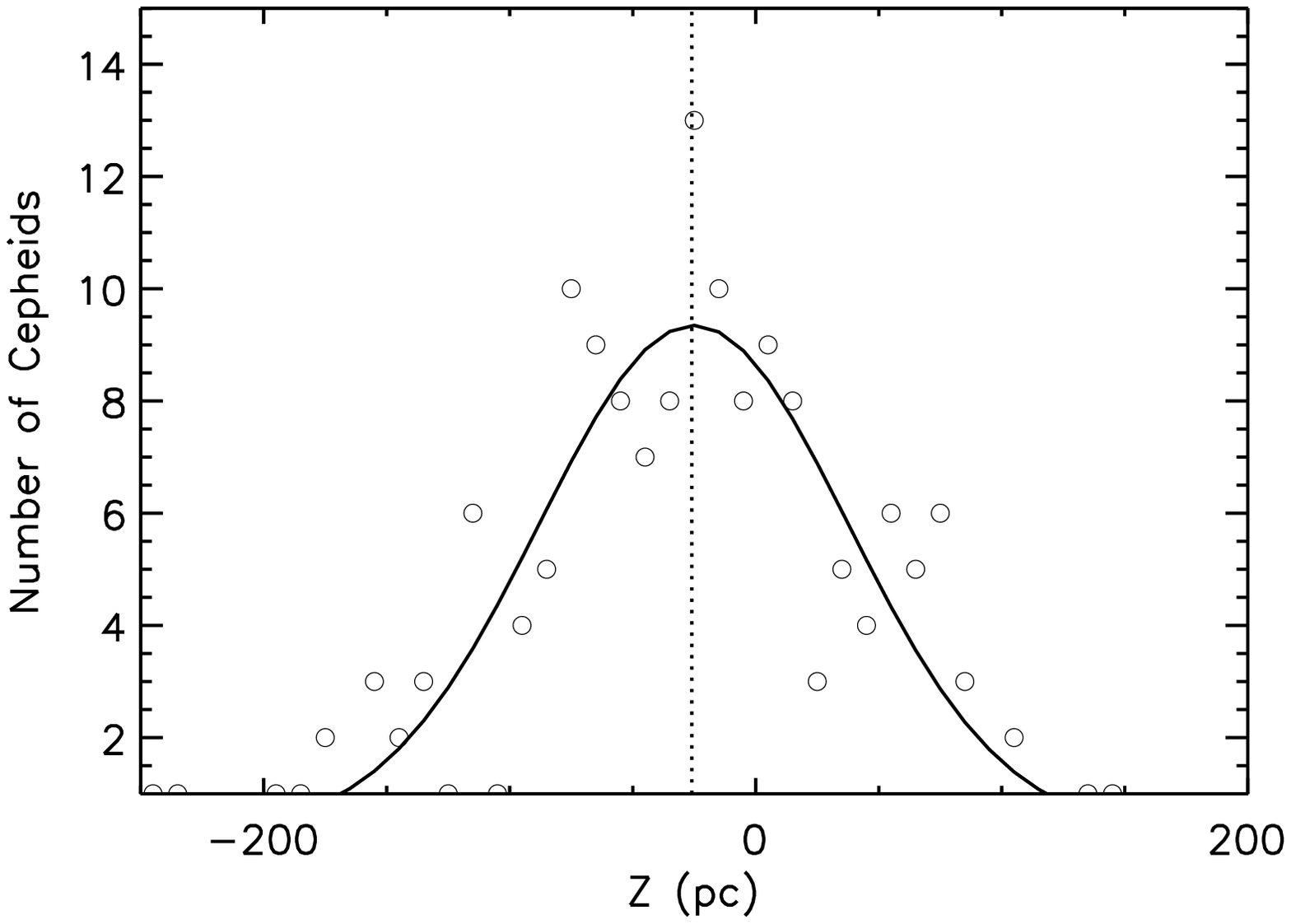}
\end{center}
\caption{Top, a plot of projected distances of classical Cepheids from the Galactic plane, as a function of distance. Note that most Cepheids lie below the plane as viewed from the Sun. Bottom, a Gaussian fit to the distribution of Cepheids for $d\le2$ kpc (binned at $\Delta Z=10$ pc). The offset in $Z$ represents the Sun's distance from the Galactic plane ($26\pm3$ pc).}
\label{fig4}
\end{figure}

\begin{figure}
\begin{center}
\includegraphics[width=8cm]{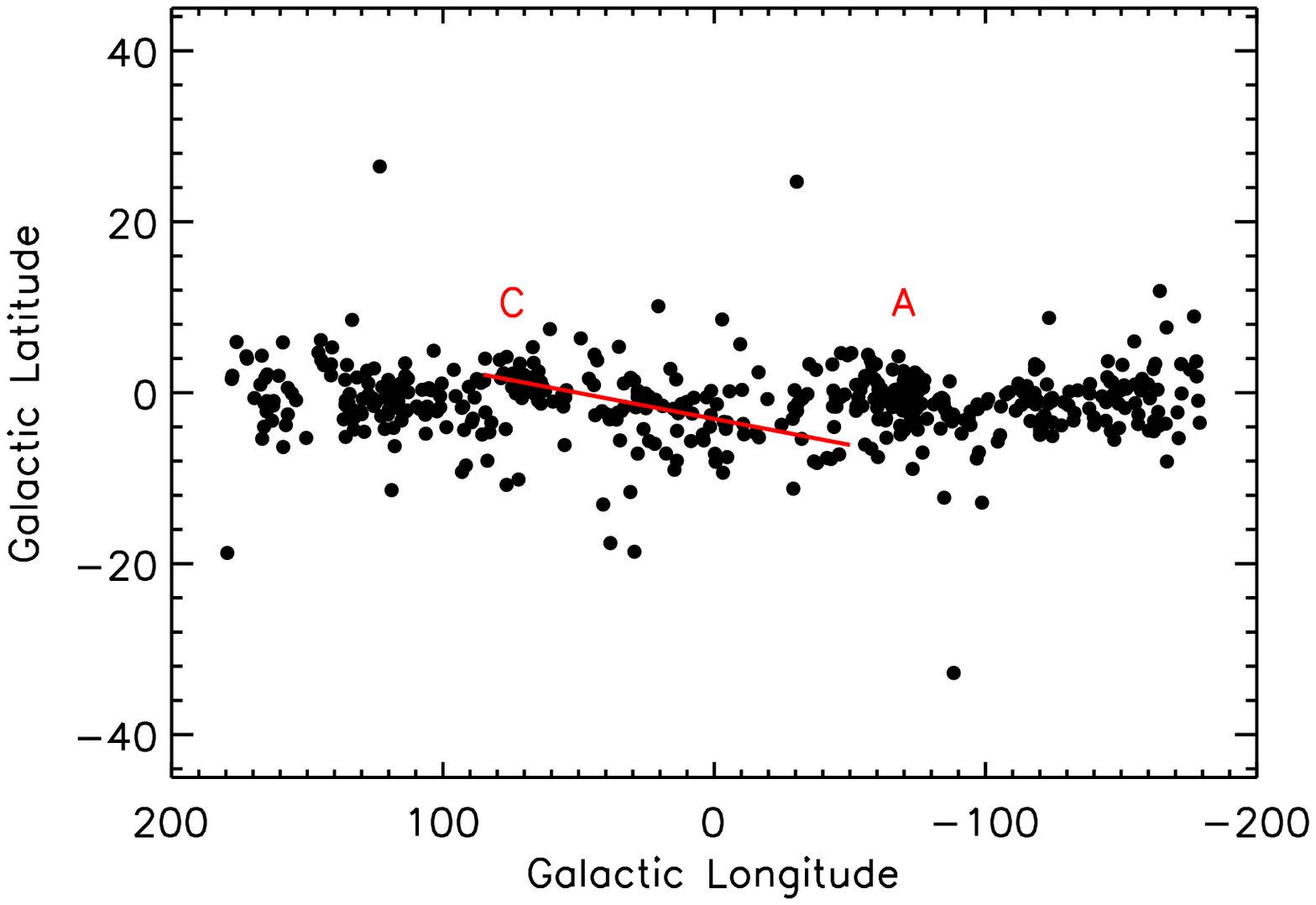}
\includegraphics[width=8cm]{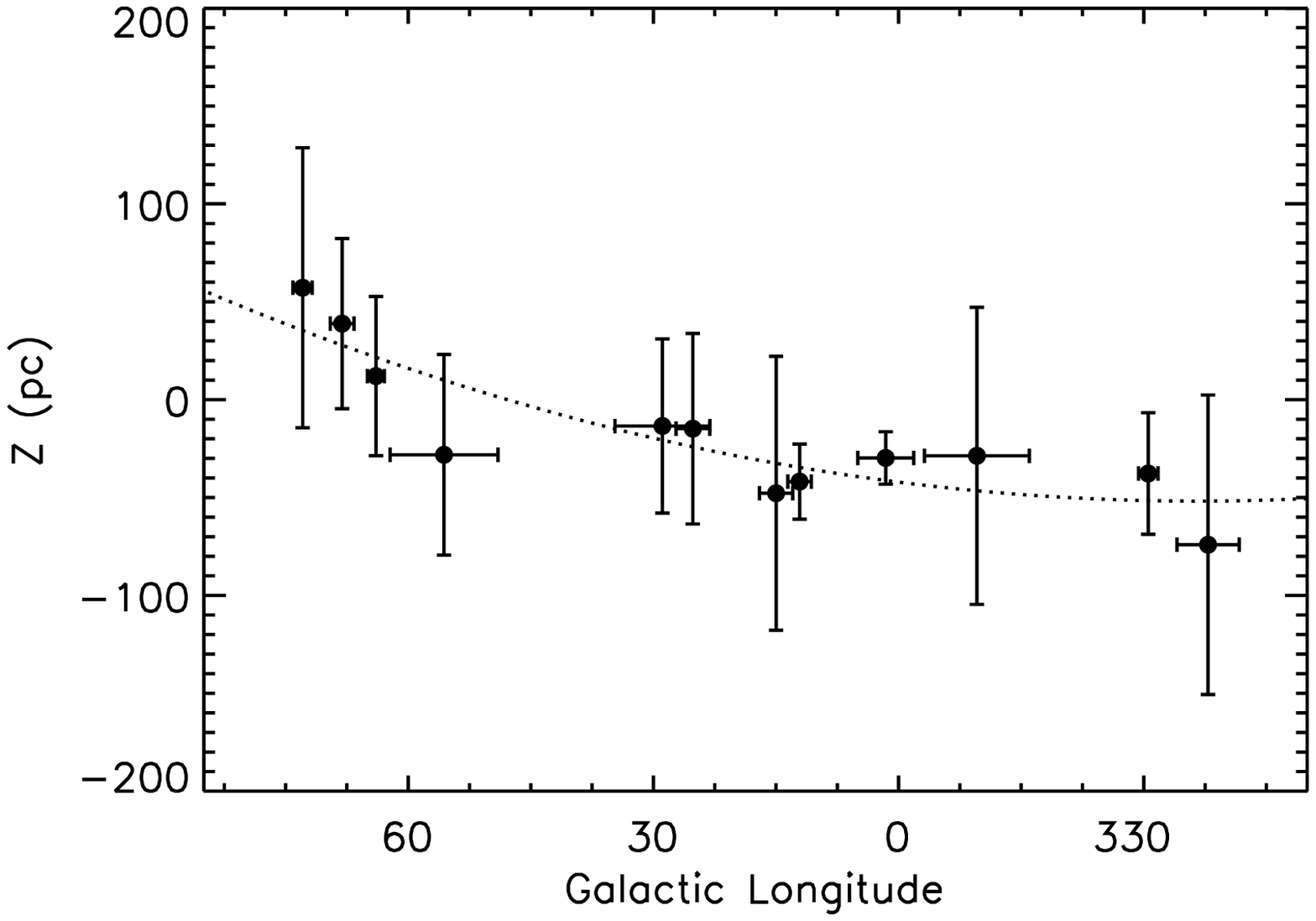}
\caption{Top, the skewed distribution of classical Cepheids. Features A and C represent directions towards the Sagittarius-Carina arm and the Cygnus feature, respectively (see \S \ref{spiralsection}).  Bottom, the Galactic longitude dependence of apparent distance from the Galactic plane for classical Cepheids.}
\label{fig5}
\label{fig6}
\end{center}
\end{figure}

A plot of projected distance from the Galactic plane for each classical Cepheid as a function of distance from the Sun is plotted in Fig. \ref{fig4} (top). The diagram is plotted relative to the view from the Sun in order to illustrate the skewed distribution of classical Cepheids from the solar perspective. The inclination of the local spiral arm, coincident with Gould's Belt, and warping of the disk (Fig. \ref{fig5}), can lead to potential bias in determining the Sun's distance from the Galactic plane. It is therefore important to select the sample for analysis as a function of distance and direction. Distant classical Cepheids in the Cygnus direction ($\ell\simeq70 \degr$), for example, appear to lie above the plane relative to distant classical Cepheids in the direction of Sagittarius (Fig. \ref{fig5}). The two regions, separated by $\sim 100$ pc in $Z$, represent extrema where the Sun's distance from the Galactic plane inferred from classical Cepheids will be either above average for the Sagittarius sample, to below for the Cygnus sample.  However, there may be a preference towards detecting classical Cepheids at larger galactic latitudes owing to increased interstellar extinction along the plane, introducing a potential bias.  Nevertheless, the signature of warping as illustrated by classical Cepheids is in general agreement with the results of \citet{lc02}, \citet{ru03}, and \citet{vi05}.

The Sun's distance from the Galactic plane can be established reasonably well from classical Cepheids in the local sample (within $d \le 2$ kpc), where the effect of the Milky Way's warp is small yet the variables are sampled beyond features associated with Gould's Belt. Such an analysis gives $Z_{\sun}=26\pm3$ pc, as determined from the offset of a Gaussian fit to the data (Fig \ref{fig4}). However, the systemic uncertainty may be larger than the formal uncertainty cited owing to the effects described above. Literature results for $Z_{\sun}$ lie between $5$ and $30$ pc, as tabulated by \citet{re97,re06} and \citet{jo05,jo07}. \citet{re06} used the distribution of OB stars to derive a value of $Z_{\sun}=19.6\pm2.1$ pc. \citet{jo05} inferred a value of $Z_{\sun}=22.8\pm3.3$ pc on the basis of interstellar extinction towards open clusters, and a more recent analysis of young open clusters and OB stars produced values of $Z_{\sun}=13$ to 20 pc and $Z_{\sun}=6$ to 18 pc, respectively \citep{jo07}. Star counts were used by \citet{hu95} to obtain a value of $Z_{\sun}=20.5\pm3.5$ pc. The present result from classical Cepheids is slightly larger than the above estimates.

\begin{figure}
\begin{center}
\includegraphics[width=8cm]{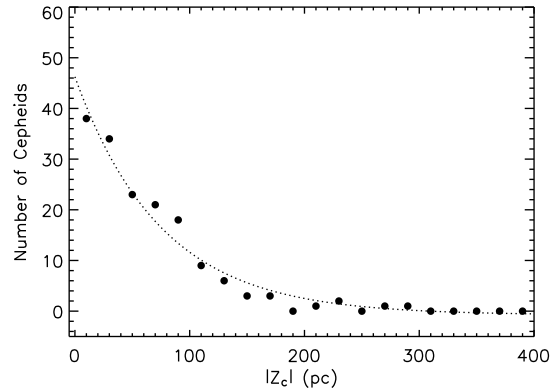}
\caption{The number of classical Cepheids, sampled in 20 pc bins as a function of $Z$, decreases with increasing distance from the Galactic plane.}
\label{fig7}
\end{center}
\end{figure}

Classical Cepheids present several advantages for such an analysis, since the distances to individual classical Cepheids can generally be estimated more precisely than the distances to individual OB stars or open clusters. Conversely, OB stars exhibit a spread in luminosity with spectral type \citep[e.g.,][]{tu76,tu79}, although the inverse is true for their intrinsic colours. The precision of distances derived for individual OB stars is therefore contingent on the availability of precise MK spectral types.

Likewise, distances to individual open clusters are often poorly constrained, for various reasons. Even among bright Messier objects (e.g., M38, M46) and calibrating Cepheid clusters, there can be unsatisfactory scatter in derived distances \citep{ma07,ma08a}. In some cases the distances to clusters derived in recent studies are nearly twice as large as values obtained previously \citep[e.g., NGC 2452,][]{ga84,ma95,mo01}, or inferences about their evolutionary ages and constituent stars are completely revised \citep[e.g., King 13,][]{ma08a}.

\begin{figure*}
\begin{center}
\includegraphics[width=16cm]{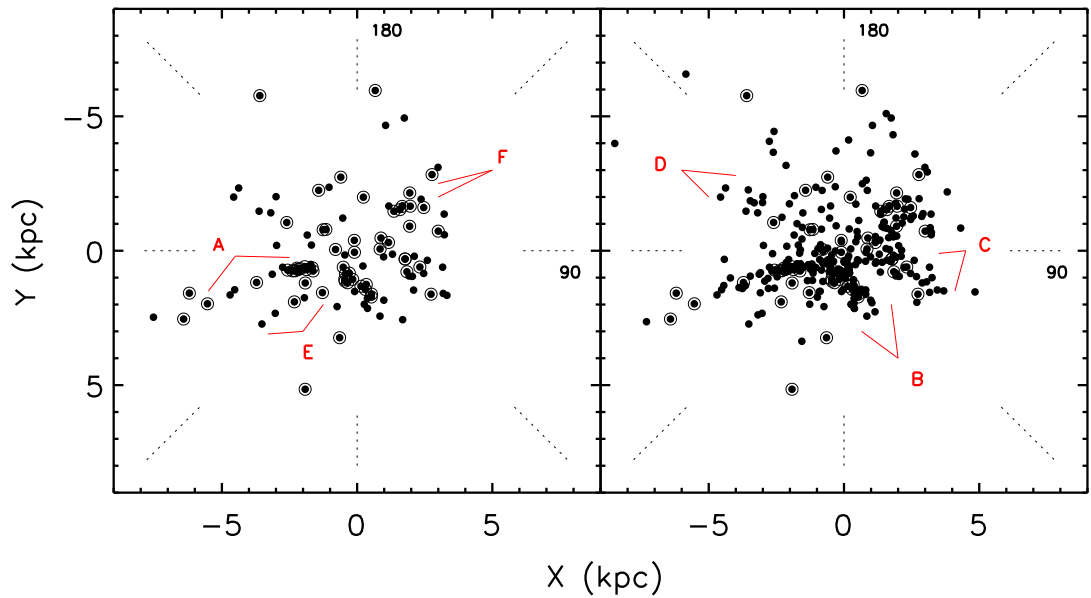}
\end{center}
\caption{Local spiral structure as delineated by classical Cepheid variables (solid points) and young open clusters (YOCs, circled points) in Galactic Cartesian space centred on the Sun ($X,Y=0$). Hybrid maps are presented for long-period Cepheids ($P \ge13$ days) and YOCs (left), and including short-period Cepheids ($P \ge5$ days, right). Markers refer to features discussed in the text.}
\label{fig8}
\end{figure*}

Classical Cepheids are sparsely distributed near the Sun, with the nearest classical Cepheid, Polaris \citep{tu05,ev08}, more than 100 pc distant and the bulk of the sample beginning to appear at distances of $\sim 250$ to 300 pc. A plot of the distribution of classical Cepheids with distance from the Galactic plane ($d \le 2$ kpc) is presented in Fig. \ref{fig7}. The data, binned to reduce the scatter, have a functional dependence given by: $\rho=47 \times e^{- |Z_c|/75}-0.76$, which implies a classical Cepheid scale height of $Z_h \le75\pm10$ pc, similar to the value of $70\pm10$ pc obtained by \citet{fe68}. $|Z_c|$ is the absolute distance of a classical Cepheid from the Galactic plane after correction for the solar bias ($Z_{\sun}=26$ pc). A further bias arises when using samples covering great distances because of the warping of the disk and interstellar extinction, discussed earlier, which artificially increases the determined scale height. The scale height derived here, and likely by other means, is therefore an upper limit.

There are 80 classical Cepheids within 1 kpc of the Sun, a number that presumably underestimates the true sample size. If that number is assumed to be typical of the rest of the Galactic disk, and the disk is assumed to populate the region between 1.2 kpc (excluding the bulge) and $\sim 13$ kpc from the Galactic centre, then the total number of classical Cepheids in the Galaxy is at least 15,000.

\section{Galactic Spiral Structure}
\label{spiralsection}
W. W. Morgan's first delineation of the spiral arms of our Galaxy using early-type stars was a highlight of the 1951 meeting of the American Astronomical Society \citep{ga95}, and marked the culmination of a century of speculation about the nature of the Milky Way. \citet{al52} appears to have been the first to argue that ``the Milky Way and the stars within it together constitute a spiral with several (it may be four) branches, and a central (probably spheroidal) cluster.'' Decades later \citet{pr69} and \citet{ea00} also wrote about the Milky Way's spiral structure, although neither referenced Alexander's visionary treatise. In many works the Sun was assumed to lie at the centre of the spiral pattern, with \citet{ea00} suggesting an alternative centre. Currently, the canonical Galactic model is that of a 4-armed grand design spiral \citep[a convenient summary is provided by][]{va05}, yet some well-established and well-populated young Galactic features are not matched by the superposed spiral patterns, and in some instances the superposed patterns pass through regions of the Galaxy devoid of spiral arm tracers. The empirical picture of spiral arms in our Galaxy appears to be problematic.

Interstellar extinction prevents a complete delineation of Galactic structure by classical Cepheid variables, limiting an analysis to the local vicinity of the Sun. Nevertheless, the analysis reveals features that both support and contradict the seminal work by \citet{ge76}, \citet{ru03}, and \citet{va08}. A plot of the distribution of classical Cepheids and young open clusters \citep[YOCs, compiled from the catalogs of][]{di02,mp03} in Cartesian space is presented in Fig. \ref{fig8}. For the present investigation, young open clusters are defined to be those with turnoff spectral types of B1 or earlier (ages $\le10^7$ years). Two separate Cepheid samples were utilized, one consisting of long-period classical Cepheids and YOCs, and a sample that also includes shorter-period classical Cepheids ($P \ge 5$ d). The spread of periods for the latter sample includes stars of lower progenitor mass, and hence older evolutionary age, than the former \citep{tu96b,tu06}. Old, low-mass stars like Type II Cepheids \citep{wa02} are obviously excluded from such an analysis.

It is generally considered that only the most massive and youngest stars are suitable for deleneating spiral structure, since they have not progressed far from their places of birth in the spiral arms. Yet the consistent picture established between long-period and short-period classical Cepheids in Fig. \ref{fig8} suggests that short-period classical Cepheids are sufficiently young to delineate spiral features as well. They have main sequence progenitors of at least 4--6 $M_{\sun}$ and correspond to ages of 40--80 Myr. Short-period classical Cepheids have therefore covered less than $\sim30\%$ of their Galactic orbits, which means they have not drifted far from their birthplaces. More importantly, long-period classical Cepheids and YOCs produce a consistent picture of the Galaxy. Use of such tracers simultaneously provides a larger statistical sample and independent confirmation of the results, inevitably providing more confident conclusions. 

The Sagittarius-Carina feature (A) is considered to be one of the Galaxy's major spiral arms, as confirmed by the distribution of classical Cepheids and YOCs. Classical Cepheids concentrate heavily along its length, traced by objects like U Car, VY Car, XZ Car, SV Vel, RY Vel, and YZ Car. But the canonical spiral pattern has the arm originating from Galactic longitudes in excess of $\ell \simeq 35\degr$, passing through a region almost devoid of optical tracers. That discrepancy was studied by \citet{fo83,fo84,fo85}, and was attributed partly to the presence of heavy extinction arising within a nearby giant molecular cloud lying in that direction, as well as to a dirth of spiral arm tracers. Preliminary data from the Abbey Ridge Observatory \citep{la07,ma08b} for three newly-discovered Cepheids \citep{wo04,wg04} lying in that general direction confirm that the extinction here is exceptionally large at nearly $A_V\simeq4$ magnitudes per kiloparsec (the photometry and relevant details shall be published in a subsequent study). Sample incompleteness may therefore be important. However, the distribution (B) in Fig. \ref{fig8} strongly suggests that the Sagittarius-Carina arm (A) originates from a different region of the Galaxy. Feature (B) appears to be outlined by classical Cepheids like AV Sgr, VY Sgr, WZ Sgr \citep{tu93}, UZ Sct, RU Sct, and Z Sct.

The Cepheid/YOC picture also indicates a feature emanating from Vulpecula-Cygnus (C), tied to variables like S Vul, AS Vul, GQ Vul, TX Cyg, CD Cyg, SZ Cyg, and VX Cyg. The feature appears to continue locally near the Sun, where it runs closely adjacent to the Sagittarius-Carina arm. The picture is rather ambiguous, however, and it is difficult to establish the existence of a continuous spiral feature running into the third Galactic quadrant.

The region surrounding the Sun is relatively complex, containing numerous young objects and a juxtaposition of several spiral features. There is a concentration (D) in the direction of the Puppis associations (e.g., Pup OB1 and Pup OB2), ranging from $\sim$3--4 kpc and tied to classical Cepheids such as EK Pup, AQ Pup, SS CMa, X Pup, WZ Pup, BN Pup, WY Pup, and WW Pup, and classical Cepheids near $\sim$5 kpc, like AD Pup and LS Pup. The picture {\it hints} at the possibility that the Puppis associations may be an extension of the local feature described above (C) or a spur of the Sagittarius-Carina arm. Examination of images of spiral galaxies in an atlas like that of \citet{sb88} indicates that galaxies with spiral arms that twist, merge, cross, divide into two, and exhibit smaller spurs in their outer regions are indeed frequent. Conversely, purely well-behaved grand design spirals are much less common.

Cepheids are concentrated in the Cassiopeia feature (F). Although, the well-known depletion of the Perseus arm for $\ell \ge 140\degr$ also shows up in the Cepheid distribution, and it is difficult to trace a major spiral feature beyond that point. Long-period Cepheids also suggest the presence of a minor spiral feature (E) that is tied to variables in Centaurus like QY Cen, KN Cen, and VW Cen. 

Matching the distribution of classical Cepheids and YOCs to a standard spiral pattern is rather challenging, so no superposition of such a pattern has been made in Fig. \ref{fig8}. The figure has been tagged, however, with several identifiers that relate to features discussed above.

\section{Summary}
A new Type II Cepheid reddening-free distance parameterization is formulated from OGLE LMC Cepheids (equation \ref{eqn2}). The {\it VI} reddening-free Type II Cepheid distance relation reproduces the calibrating set with an average uncertainty of $\sim5$\%. The distances to individual Type II Cepheids are estimated to be no larger than 5--15\%.  The median distance computed to a sample of Type II Cepheids lying in the direction of the bulge yields a distance to Galactic centre of $R_0 = 7.8\pm0.6$ kpc, with the caveat that the Type II Cepheids are assumed to be symmetrically distributed about the latter. A second estimate was established by adding an estimate for the radius of the Galactic bulge ($\beta$) to the distance to its near side ($R_{NS}$) as identified by Type II Cepheids, yielding $R_0 = R_{\rm NS}+\beta=7.7\pm0.7$ kpc.  The resulting estimates for $R_0$ from the {\it VI} reddening-free Type II Cepheid distance relation agree closely with literature values. The true uncertainties in our estimated distances to the Galactic centre may be larger than the standard errors cited, however, given that the sample of bulge Type II Cepheids is small, exhibits much scatter, and a potential metallicity effect cannot be excluded.  There is also an apparent dependence of distance with pulsation period for bulge Type II Cepheids, a trend not observed in Type II Cepheids belonging to the metal-rich globular cluster NGC 6441. It is noted that \citet{ud03} discovered large variations in the extinction law towards the bulge, which may complicate matters. The robustness of the {\it VI} reddening-free Type II Cepheid distance relation was tested using independent samples of Type II Cepheids in globular clusters and galaxies.  The distances computed to Type II Cepheids in the globular clusters M54, M92, NGC 6441, M5, and M15 by means of equation \ref{eqn2} agree with estimates found in the literature. The globular clusters exhibit a large range in metallicity ($\Delta [Fe/H]\simeq1.75$, \citet{ha96}), so the close agreement of the present distance estimates with literature results allays concerns regarding a sizeable metallicity effect. Type II Cepheids are also confirmed as likely members of the galaxies NGC 3198 and NGC 5128, respectively, once their distances are computed with the appropriate parameterization (equation \ref{eqn2}). The variable in NGC 3198 may be the most distant Type II Cepheid established to date, with an estimated distance of $d=13.7\pm3.6$ Mpc. The uncertainties are large, however.  Yet such cases demonstrate the potential use of Type II Cepheids for extragalactic research and as yet another means for testing the dependence of metallicity on Cepheid distance determinations.

The maximum thickness of the bulge along $\ell\simeq 0 \degr$ is estimated to be $H_B=2.5\pm0.3$ kpc from bulge planetary nebulae and an adopted distance to the Galactic centre. 

The Sun's distance above the plane is inferred from classical Cepheids to be $Z_{\sun}\simeq26\pm3$ pc. The determination is hampered by local effects arising from Gould's Belt and warping in the disk, requiring prudence in selecting a subsample for analysis which is representative of the region near the Sun. The signatures of Gould's Belt and the Galactic warp are evident from distant classical Cepheids in the Cygnus direction ($\ell\simeq70 \degr$) appearing to lie well above the plane relative to distant classical Cepheids located in the direction of Sagittarius. The two clumps of Cepheids are separated by $\simeq 100$ pc in $Z$.   A potential bias may arise because of a preference towards detecting classical Cepheids at larger galactic latitudes owing to increased interstellar extinction along the plane. The classical Cepheid scale height is estimated to be $Z_h \le75\pm10$ pc, a value cited as an upper limit because of the bias imposed by the disk's warp and interstellar extinction, which can artificially increase the derived result. The aforementioned bias likely affects the determination of the scale height by other means. The total number of classical Cepheids in the Galaxy is estimated to be on the order of $\sim 15,000$. 

Cepheid variables and young open clusters concentrate in obvious and consistent patterns typical of local spiral arms.  The inferred picture of such features both supports and contradicts existing interpretations. The Sagittarius-Carina arm is confirmed as a major spiral arm that appears to originate from a different Galactic region than suggested previously. A major feature is also concentrated in Cygnus-Vulpecula and may continue locally near the Sun into the third quadrant, possibly extending into the Puppis associations. More work is needed to complete the picture, however. Short-period classical Cepheids are shown to be as useful as long-period Cepheids as tracers, indicating that stars born in spiral arms remain close to their places of origin for at least $\sim80$ Myr.

The future {\it GAIA} mission \citep{ct06}, a next generation follow-up to the {\it Hipparcos} mission, should detect a large sample of new Cepheids that may help to elucidate the Milky Way's structure, in addition to the discoveries of new Galactic open clusters \citep{di02,al03,mo03,kro06,bo08,tu09}. Indeed, a multifaceted approach will likely be needed to clarify the presently available evidence pertaining to the Sun's location relative to the main components of the Galaxy. The present study appears to support the historic tradition of utilizing Cepheid variables in such an endeavour. 

\subsection*{ACKNOWLEDGEMENTS}
We are indebted to Leonid Berdnikov, Laszlo Szabados, and the staff of OGLE, whose comprehensive work on Cepheid variables was invaluable to our analysis, Doug Welch (McMaster Cepheid Archive), Michael Sallman (TASS), Grzegorz Pojmanski (ASAS), Arne Henden and Michael Saladyga (AAVSO), Brent Miszalski (MASH), Alison Doane (HCO), Carolyn Stern Grant (ADS), and the folks at CDS. Reviews by \citet{fe02}, \citet{fe99,fe01}, Anton Pannekoek, and David Higgins, were useful in the preparation of this work. Lastly, a special thanks is reserved for Charles Bonatto, George Jacoby, Noel Carboni, Petr Pravec, Joris Van Bever, Jerry Bonnell, Jason Kalirai, David Balam, David Bartlett, Petr Pravec, Jerry Bonnell, Jay Anderson and the RASC.

\end{document}